\documentclass[11pt,oneside]{article}

\usepackage{amsmath}
\usepackage{graphicx}

\begin{document}
\titlepage

\title{Stimulated Raman Transfer of SIT Solitons}
\author{B.D. Clader and J.H. Eberly \\
\small{Department of Physics and Astronomy} \\
\small{University of Rochester} \\
\small{Rochester, NY 14627}}
\date{}

\maketitle

\begin{abstract}
We treat theoretically the two-pulse plane-wave propagation problem associated with the stimulated Raman process. A strong pump pulse and a weak probe pulse interact self-consistently during mutual propagation through a medium in the three-level lambda configuration. We treat the pulse interaction fully coherently, but take initially incoherent  mixed ground states to describe the medium.  We present analytic pulse envelope solutions to the nonlinear evolution equations, and then test them numerically.   We show that the entire process can be interpreted consistently as a three-regime evolution, in which the first and last regimes display behavior quantitatively the same as self-induced transparency. They are connected through a Raman transfer process in the intermediate regime.
\end{abstract}

\section{Introduction}
It is a pleasure to join the other authors of this special volume in celebrating the exceptionally wide variety of accomplishments in the distinguished career of Professor Girish S. Agarwal. His impressive contributions as a theoretical physicist in quantum optics began with more than two dozen papers written in our Department in Rochester while still a student \cite{agarwal0}. It seems certain we will continue to benefit from many further interesting calculations and valuable insights in years to come.

As coherence in quantum optical contexts has provided a constant source of fruitful inspiration to Professor Agarwal, we've chosen the Raman effect during fully coherent pulse propagation in three-level media as the topic of our contribution. This is an area that has prompted recent studies of his own \cite{vemuri-etal,martinezL-agarwal, agarwal-eberly, panigrahi-agarwal, agarwal-dey, agarwal-etal}, and is an area of wide theoretical and experimental activity at present. The door to this domain was opened by the famous work of McCall and Hahn \cite{mccall-hahn} on two-level effects in ruby, in which they found theoretically and demonstrated experimentally the first optical soliton, the $2\pi$ $sech$ pulse of self-induced transparency (SIT).

The coherent pulse propagation domain was gradually expanded to include pulse-pair propagation including three-level solitonic behavior \cite{vemuri-etal, konopnicki-eberly, bolshov-etal, eberly, park-shin}, and the phenomenon of electromagnetically induced transparency (EIT) then emerged in the seminal work of Harris \cite{harris}. This has led to exciting discoveries including superluminal and ultra-slow pulse propagation. A consistent theoretical challenge, particularly in numerical modeling of short pulses, has been the effect of inhomogeneous broadening (see some treatments in \cite{agarwal-dey, konopnicki-eberly, eberly-etal, clader-eberly}), and that is one element of the present paper.

The Raman effect was the first effect in physics of a specifically three-level character, and it is the basic model for essentially all experiments on lambda systems with two optical fields. Our contribution can be considered as an examination of the {\em cooperation} between SIT and the Raman process, in contrast to the study by Agarwal, Dey and Gauthier \cite{agarwal-etal} on the {\em competition} between EIT and Raman processes. Of course the stimulated Raman effect (STR) has been actively studied for decades (for an early review see \cite{bloembergen}).  The STR is a process by which a strong laser beam incident on the ``pump" transition of a three-level lambda system causes amplification of a weak ``probe" pulse injected at the Stokes frequency (see Fig. 1).  Traditionally STR has been studied in atomic and molecular systems, but it has also been examined in Bose condensates \cite{martinezL-agarwal}.

In this paper we present analytic and numerical solutions to the coupled plane-wave Maxwell-Bloch system for stimulated Raman scattering, without the usual assumption of undepleted pump pulse and with careful consideration of non-trivial changes in both the probe and pump pulse shapes due to propagation.  We show in the fully coherent case (pulses short enough to neglect homogeneous damping) that complicated exact analytical results can be interpreted relatively easily, partly in SIT terms, by introducing three ``regimes" of interaction. We show how the three regimes can be considered separately despite the tightly coupled nonlinear nature of the solutions.

\section{Theoretical Raman Model}

\begin{figure}[h]
\begin{center}
\leavevmode
\includegraphics[width = 1in]{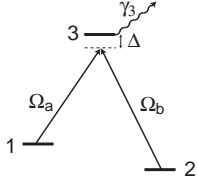}
\end{center}
\caption{\label{lambda} \small{Three-level atom in two-photon resonance with level $1$ connected to level $3$ via the Rabi frequency $\Omega_{a}$ and level $2$ connected to level $3$ via the Rabi frequency $\Omega_{b}$.  Each laser field is detuned from the excited state resonance by an equal amount $\Delta$, and the excited state decay rate is given by $\gamma_3$.}}
\end{figure}

We consider the coherent propagation of two laser pulses in the $x$ direction in a medium of three level atoms in the lambda configuration as shown in Fig. \ref{lambda}.  The electric field of the individual laser pulses can be written as $E_a(x,t) = \mathcal{E}_a(x,t)e^{-i(k_a x - \omega_a t)} + \rm{ c.c.}$, where $\mathcal{E}_a$ is the slowly varying envelope of the electric field, $k_a$ is the wavenumber, and $\omega_a$ is the frequency (similarly for $E_b(x,t)$).

The Hamiltonian of the system in the rotating wave picture is given by
\begin{equation}
\label{Hamiltonian}
H=-\hbar\frac{\Omega_{a}}{2}|1\rangle\langle 3| - \hbar\frac{\Omega_{b}}{2}|2\rangle\langle 3| + \textnormal{ h.c.} + \hbar\Delta |3\rangle\langle 3|,
\end{equation}
where h.c. refers to the hermitian conjugate of the previous operator. For the 1-3 transition we have $\Omega_a = 2d_a\mathcal{E}_a/\hbar$, $d_a$ is the dipole moment, and $\Delta = \omega_3 - (\omega_1 + \omega_a)$, and corresponding notation for the 2-3 transition. Note that the same $\Delta$ is the one-photon detuning of each transition below resonance.

We will describe each individual atom by its density matrix.  The von Neumann equation for the density matrix $i\hbar\frac{\partial \rho}{\partial t} = [H,\rho]$ gives the following density matrix equations, with the use of the Hamiltonian in Eq. \eqref{Hamiltonian}:

\begin{subequations}
\label{DensityMatrix}
\begin{align} \dot{\rho}_{11} & = i \frac{\Omega_{a}}{2}\rho_{31} - i\frac{\Omega_{a}^*}{2}\rho_{13} \\
\dot{\rho}_{22} & = i \frac{\Omega_{b}}{2}\rho_{32} - i\frac{\Omega_{b}^*}{2}\rho_{23} \\
\dot{\rho}_{33} & = -i \frac{\Omega_{a}}{2}\rho_{31} + i\frac{\Omega_{a}^*}{2}\rho_{13} - i \frac{\Omega_{b}}{2}\rho_{32} + i\frac{\Omega_{b}^*}{2}\rho_{23} \\
\dot{\rho}_{12} & = i \frac{\Omega_{a}}{2}\rho_{32} - i \frac{\Omega_{b}^*}{2}\rho_{13} \\
\dot{\rho}_{13} & = i \frac{\Omega_{a}}{2}(\rho_{33}-\rho_{11}) - i \frac{\Omega_{b}}{2}\rho_{12} + i\Delta \rho_{13} \\
\dot{\rho}_{23} & = i \frac{\Omega_{b}}{2}(\rho_{33}-\rho_{22}) - i \frac{\Omega_{a}}{2}\rho_{21} + i\Delta \rho_{23}.
\end{align}
\end{subequations}
Eqns. \eqref{DensityMatrix} assume that the duration of the laser pulses is sufficiently short that the excited state decay rate $\gamma_3$ can be neglected because we consider pulses with durations much shorter than $1/\gamma_3$.  Maxwell's equation for the field gives these two slowly varying envelope equations
\begin{subequations}
\label{MaxwellEquation}
\begin{align}
\left( c \frac{\partial}{\partial x} + \frac{\partial}{\partial t}\right) \Omega_{a} & = -i\mu_{a}\langle\rho_{13}\rangle \\
\left( c \frac{\partial}{\partial x} + \frac{\partial}{\partial t}\right) \Omega_{b} & = -i\mu_{b}\langle\rho_{23}\rangle,
\end{align}
\end{subequations}
where $\mu_a=Nd_a\omega_a^2/\hbar\epsilon_0$ is the atom field coupling parameter, $N$ is the density of atoms, $d_a$ is the dipole moment matrix element between atomic states $1 \to 3$, and similarly $\mu_b$ for states $2 \to 3$.  The notation $\langle \hdots \rangle = \int_{-\infty}^{\infty}(\hdots)F(\Delta)d\Delta$ accounts for inhomogeneous broadening by averaging over a range of detunings $\Delta$ with the function $F(\Delta) = (T_2^*/\sqrt{2\pi}) e^{-(T_2^*\Delta)^2/2}$, where $T_2^*$ is the inhomogenous lifetime. For these equations to permit analytic solutions, one needs $\mu_{a}=\mu_{b} \equiv \mu$, which we will assume for the remainder of this paper.

\section{Pulse Propagation Analytic Solutions}

We solve Eqns. \eqref{DensityMatrix} and \eqref{MaxwellEquation} using a method developed by Park and Shin \cite{park-shin} which is based on the B\"acklund transformation.  We assume the initial density matrix of each atom is in an incoherent mixture of the two ground states, which is written in explicit form as
\begin{equation}
\label{InCondition}
\rho(0) = \begin{pmatrix}
|\alpha|^2 & 0 & 0 \\
0 & |\beta|^2 & 0 \\
0 & 0 & 0
\end{pmatrix}.
\end{equation}
The solution to the field envelopes is given by
\begin{subequations}
\label{PulseSolution}
\begin{align}
\Omega_{a} & = \frac{4}{\tau}\bigg[2 \textnormal{ cosh}\big(T/\tau -Ax\big) + \exp\big(T/\tau +(A-2B)x\big)\bigg]^{-1} \\
\Omega_{b} & = \frac{4}{\tau}\bigg[2 \textnormal{ cosh}\big(T/\tau -Bx\big) + \exp\big(T/\tau +(B-2A)x\big)\bigg]^{-1},
\end{align}
\end{subequations}
where $\tau$ is the nominal pulse width and $T=t-x/c$ is the retarded time, and $A$ and $B$ are given by
\begin{subequations}
\label{lengthScale}
\begin{align}
A & = |\alpha|^2\frac{g(\tau,T_2^*)}{2} \\
B & = |\beta|^2\frac{g(\tau,T_2^*)}{2},
\end{align}
\end{subequations}
where the function $g(\tau,T_2^*)$ is the Beers absorption coefficient (inverse Beers length) sometimes written $\alpha_D$ in the case of Doppler broadening:
\begin{eqnarray}
\label{BeersLength}
g(\tau,T_2^*) = \frac{\mu}{\tau c} \int_{-\infty}^{\infty}\frac{F(\Delta)d\Delta}{\Delta^2 + \left(\frac{1}{\tau}\right)^2} \equiv \alpha_D.
\end{eqnarray}

Related solutions were previously given in \cite{bolshov-etal} and \cite{park-shin} for the case where $|\alpha|^2=1$, and in the limit where inhomogeneous broadening is ignored.  For the remainder of this paper we will be considering the opposite limit, corresponding to large inhomogeneous broadening as in the familiar SIT scenario \cite{mccall-hahn}, and $\alpha_D^{-1}$ will be the appropriate length unit for spatial pulse evolution.
\begin{figure}[h]
\begin{center}
\leavevmode
\includegraphics{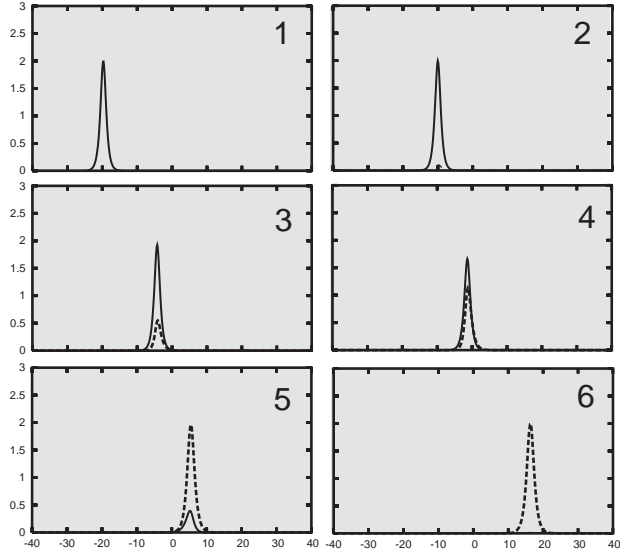}
\end{center}
\caption{\label{fig.pulse} \small{Plots of the analytic pulse solutions given in Eq. \eqref{PulseSolution}.  The horizontal axis is $x$ in units of the absorption length $\alpha_D^{-1}$, and the vertical axis is the pulse Rabi frequency in units of $\tau^{-1}$. The background is slightly shaded to indicate the presence of the lambda medium.  The solid curve is $\Omega_{a}$ and the dashed curve is $\Omega_{b}$.  The plot shows Raman amplification as an exchange process between two SIT solitons.  The input pulse on the 1 - 3 transition amplifies a weak probe pulse on the 2 - 3 transition. Parameters: $|\alpha|^2 = 0.8$, $|\beta|^2=0.2$, $\tau \approx 3T_2^*$.}}
\end{figure}

\section{Three-Regime Behavior}

The analytic pulse solutions are plotted in Fig. \ref{fig.pulse}. Their solution formulas \eqref{PulseSolution} resist easy interpretation.  However, they can be unravelled as follows.  From the plots we see three distinct types of evolution.  When only $\Omega_{a}$ is significantly present  we will speak of regime I (frames 1 and 2 of Fig. \ref{fig.pulse}).  During ``transfer" from pump to Stokes when both pulses are substantial we have regime II (frames 3-5).  During regime III only $\Omega_{b}$ is significantly present (frame 6).  We will isolate these three regimes analytically by looking at the asymptotic behavior of the pulse solutions.By considering $x \ll -\alpha_D^{-1}$ we can look at regime I, and by considering $ x \gg \alpha_D^{-1}$ we can look at regime III.  The transfer regime II corresponds to $x \approx 0$.

The Area of a pulse is defined to be
\begin{equation}
\label{pulseArea}
\theta(x) = \int_{-\infty}^{\infty}\Omega(x,t)dt,
\end{equation}
from which the pulse Areas of solutions \eqref{PulseSolution} can be shown to be
\begin{subequations}
\label{PulseArea}
\begin{align}
\theta_{a}(x) & = \frac{4}{h(x)}\big[\tan^{-1}(h(x))+\cot^{-1}(h(x))\big] \\
\theta_{b}(x) & = \frac{4}{h(-x)}\big[\tan^{-1}(h(-x))+\cot^{-1}(h(-x))\big]
\end{align}
\end{subequations}
where
\begin{equation}
\label{AreaFunction}
h(x) = (1+e^{(A-B)x})^{1/2}.
\end{equation}
In the asymptotic limit $x \ll -\alpha_D^{-1}$ it is straightforward to show that $\theta_{a}(x) \approx 2\pi$ and $\theta_{b}(x) \approx 0$.  In the opposite limit, where $x \gg \alpha_D^{-1}$, we have the opposite result, $\theta_{a}(x) \approx 0$ and $\theta_{b}(x) \approx 2\pi$.

In regime I, Eq. \eqref{PulseSolution} can be approximated by (recalling the definitions of $A$, $B$, and $T$):
\begin{subequations}
\label{PulseInput}
\begin{align}
\Omega_{a} & \approx \frac{2}{\tau}\textnormal{ sech }\left(\frac{t-x/c}{\tau}-|\alpha|^2 \frac{\alpha_D}{2}x\right) \\
\Omega_{b} & \approx 0,
\end{align}
\end{subequations}
which corresponds to a single $2\pi$ sech-shaped pulse traveling at a group velocity of $v_g = c/(1+Ac\tau)$.  In regime III the pulse solutions are approximated by
\begin{subequations}
\label{PulseOutput}
\begin{align}
\Omega_{a} & \approx 0 \\
\Omega_{b} & \approx \frac{2}{\tau}\textnormal{ sech }\left(\frac{t-x/c}{\tau}-|\beta|^2 \frac{\alpha_D}{2}x\right),
\end{align}
\end{subequations}
which corresponds to a single $2\pi$ sech-shaped pulse traveling at the group velocity of $v_g = c/(1+Bc\tau)$.  Thus in both regimes I and III we see that the solution reduces to that of a $2\pi$ Area sech shaped pulse moving in a two-level medium.  These asymptotic solutions correspond exactly to SIT solitons \cite{mccall-hahn}. These two regimes are connected via the intermediate regime II where the three level character of the medium causes the Raman transfer process to occur, and the full solutions given in Eq. \eqref{PulseSolution} must be used.

\section{Numerical Solutions}

To test the potential experimental relevance of the analytic solutions presented, we have to replace the infinite uniform medium assumed to this point by a more realistic medium with finite edges. We then solve Eqns. \eqref{DensityMatrix} and \eqref{MaxwellEquation} numerically.  The analytic solution shapes and areas should not be assumed in advance, so at the entrance face to the medium we will inject gaussian input pulses given by
\begin{subequations}
\label{numInput}
\begin{align}
\Omega_{a}^{(in)} & = \frac{\theta_{a}}{\tau\sqrt{2\pi}}e^{-\frac{T^2}{2\tau^2}} \\
\Omega_{b}^{(in)} & = \frac{\theta_{b}}{\tau\sqrt{2\pi}}e^{-\frac{T^2}{2\tau^2}},
\end{align}
\end{subequations}
where the parameters $\theta_a$ and $\theta_{b}$ give the input pulse areas.  In the previous section we showed how the analytic solution can be separately interpreted in three distinct regimes by considering asymptotic solutions.  We now chose input pulses for the numerical solution specifically to test the validity of our hypothesis that this process can be broken up into the three regimes defined above.

\begin{figure}[h]
\begin{center}
\leavevmode
\includegraphics{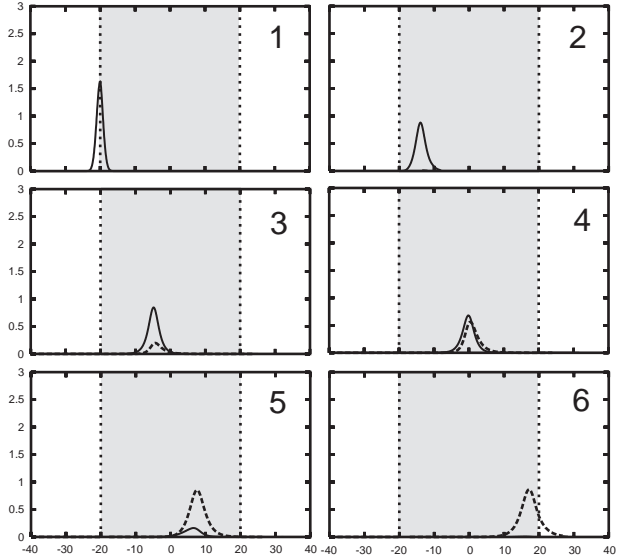}
\end{center}
\caption{\label{fig.Numpulse} \small{Plots of the numerical pulse solutions of Eqs. \eqref{DensityMatrix} and \eqref{MaxwellEquation}. The horizontal axis is $x$ in units of the absorption length $\alpha_D^{-1}$, and the vertical axis is the pulse Rabi frequency in units of $\tau^{-1}$.  The shaded zone indicates the location of the medium.  The solid curve is $\Omega_{a}$ and the dashed curve is $\Omega_{b}$.  The plot shows three distinct processes.  A two level SIT process where the input $1.3\pi$ gaussian pulse is reshaped into a $2\pi$ sech shaped pulse (frames 1-2), followed by a Raman amplification process where the input pulse on the 1 - 3 transition amplifies a weak probe pulse on the 2 - 3 transition (frames 3-5), and finally a second two level SIT process where the weak probe pulse has become a $2\pi$ sech shaped pulse (frame 6). Parameters: $|\alpha|^2 = 0.8$, $|\beta|^2=0.2$, $\tau \approx 3T_2^*$, $\theta_{a}=1.3\pi$, $\theta_{b} = 0.005\pi$.}}
\end{figure}

\begin{figure}[h]
\begin{center}
\leavevmode
\includegraphics{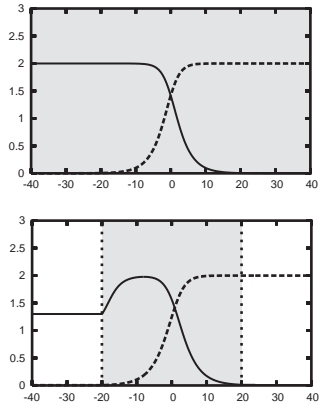}
\end{center}
\caption{\label{fig.area} \small{Plots of the pulse Areas from Eqs. \eqref{PulseArea} in the top frame, and from integrated pulse Area from the numerical solution in the bottom frame.  The horizontal axis is $x$ in units of the absorption length $\alpha_D^{-1}$, and the vertical axis is the pulse Area in units of $\pi$.  The shaded zone indicates the location of the medium.  The solid curve is $\theta_{a}(x)/\pi$ and the dashed curve is $\theta_{b}(x)/\pi$.  The top plot shows the $2\pi$ Area input pulse on the 1 to 3 transition, converted to a $2\pi$ Area output pulse on the 2 to 3 transition. The bottom curve show the input $1.3\pi$ Area pulse first converted to a $2\pi$ Area pulse consistent with the Area theorem.  The transfer process then occurs as predicted by the analytic solution.}}
\end{figure}

In Fig. \ref{fig.Numpulse} we plot the numerical solutions to Eqs. \eqref{DensityMatrix} and \eqref{MaxwellEquation}.  We assume the medium is initially configured just as in Fig. \ref{fig.pulse}, with $|\alpha|^2 = 0.8$ and $|\beta|^2 = 0.2$, and that the medium is inhomogeneously broadening such that $\tau \approx 3T_2^*$.  We choose input pulse Areas $\theta_{a} = 1.3\pi$ and $\theta_{b} = 0.005\pi$. The numerical solution shows that the $1.3\pi$ gaussian-shaped input pulse is reshaped into a $2\pi$ sech-shaped pulse in regime I just as a pure two level SIT description would suggest.  The Raman transfer process then takes place in regime II, which amplifies the weak $0.005\pi$ Stokes pulse to a $2\pi$ sech shaped pulse on the 2 to 3 transition.  This final pulse in regime III is an SIT soliton, and will propagate without any dispersion or further modification.  We also plot the numerically integrated pulse Area in the bottom frame of Fig. \ref{fig.area}.  One can clearly see the input pulse as it is reshaped into a sech shape is also changed to $2\pi$ total Area, just as predicted by standard SIT theory.

The Area of the input pulses used in the numerical solutions is an experimentally controllable parameter.  This value determines how many absorption lengths into the medium the pulses will travel in regime I before beginnning the transfer process in regime II.  The top frame of Fig. \ref{fig.area} shows the pulse Areas given in Eqns. \eqref{PulseArea}.  The input area of the weak pulse determines how far to the left of the asymptotes in Fig. \ref{fig.area} we start. In the numerical solution we choose an input pulse area for pulse b of $\theta_{b} = 0.005\pi$ which corresponds to $\alpha_Dx \approx -20$ in the analytic Area formula given in Eq. \eqref{PulseArea}. Thus within about 20 absorption lengths one would estimate the two pulses would be of roughly equal Area.  In Fig. \ref{fig.Numpulse} we indeed see the pulses in frame 3 are almost of equal Area, after traveling about 20 absorption lengths through the medium.  This can be more clearly seen in the plots of the pulse Areas in Fig. \ref{fig.area} where we see the pulses have equal Area near $\alpha_Dx \approx 20$ as predicted.  Thus we can use Eqns. \eqref{PulseArea} along with the input Area of the Stokes pulse to determine where the Raman transfer will occur even when the input pulses are of different shapes.

\section{Conclusions}
We have presented an analytic solution to the coupled Maxwell-Bloch equations that describe a Raman transfer process between two SIT solitons.  This process can be broken up into three distinct regimes of which only one exhibits the characteristics of a three-level medium.  In both asymptotic regimes the process can be considered as a two-level SIT process.  We also show how to estimate the distance that the pulses will propagate in each regime.  By knowing the input Areas of the Stokes pulses, one can use the Area formulas in Eqns. \eqref{PulseArea} to determine the propagation distance the pump pulse will propagate in regime I, as an SIT soliton.  The analytic pulse solutions in Eqns. \eqref{PulseSolution} then describe the Raman transfer process in regime II.  Finally, once the transfer process has occurred, we show that the Stokes pulse travels as an SIT soliton in regime III for the remainder of its propagation through the medium.

\section{Acknowledgements}
We thank Q.H. Park and P.W. Milonni for helpful discussions and correspondence.  B.D. Clader acknowledges receipt of a Frank Horton Fellowship from the Laboratory for Laser Energetics, University of Rochester. Research supported by NSF Grant PHY 0456952.  The e-mail contact address is: dclader@pas.rochester.edu.


\newpage

\begin{thebibliography}{99}

\bibitem{agarwal0} One record of Professor Agarwal's early interests and insightful approach to fundamental issues is contained in his account of radiative relaxation in simple model systems of atoms in {\em Quantum Statistical Theories of Spontaneous Emission and their Relation to other Approaches} (Springer-Verlag, Berlin, 1974). This little volume, still profitably consulted worldwide, is unusually comprehensive in treating radiative relaxation not only by including a number of theoretical methods, but also by dealing with both quantum and non-quantum radiation theories.

\bibitem{vemuri-etal} G. Vemuri, K.V. Vasavada, G.S. Agarwal, and Q. Zhang, Phys. Rev. A \textbf{54}, 3394 (1996).

\bibitem{martinezL-agarwal} J. Mar\'tinez-Linares and G.S. Agarwal, Phys. Rev. A \textbf{57}, 2931 (1998).

\bibitem{agarwal-eberly} G.S. Agarwal and J.H. Eberly, Phys. Rev. A {\bf 61}, 013404 (2000).

\bibitem{panigrahi-agarwal} P.K. Panigrahi and G.S. Agarwal, Phys. Rev. A {\bf 67}, 033817 (2003).

\bibitem{agarwal-dey} G.S. Agarwal and T.N. Dey, Phys. Rev. A \textbf{73}, 043809 (1996).

\bibitem{agarwal-etal} G.S. Agarwal, T.N. Dey, and D.J. Gauthier, Phys. Rev. A \textbf{74}, 043805 (2006).

\bibitem{mccall-hahn} S.L. McCall and E.L. Hahn, Phys. Rev. \textbf{183}, 457 (1969).

\bibitem{konopnicki-eberly} M.J. Konopnicki and J.H. Eberly, Phys. Rev. A {\bf 24}, 2567 (1981).

\bibitem{bolshov-etal} L.A. Bol'shov, N.N. Elkin, V.V. Likhanskii, and M.I. Persiantsev, Zh. Eksp. Teor. Fiz. \textbf{88}, 47 (1985) [Sov. Phys. JETP \textbf{61}, 27 (1985)].

\bibitem{eberly} J.H. Eberly, Quantum Semicl. Optics {\bf 7}, 373 (1995).

\bibitem{park-shin} Q-Han Park and H.J. Shin, Phys. Rev. A \textbf{57}, 4643 (1998).

\bibitem{harris} S.E. Harris, Phys. Today \textbf{50}, 36 (1997).

\bibitem{eberly-etal} J.H. Eberly, A. Rahman and R. Grobe, Phys. Rev. Letters {\bf 76}, 3687 (1996).

\bibitem{clader-eberly} B.D. Clader and J.H. Eberly, Optics Lett. {\bf 31}, 3921 (2006).

\bibitem{bloembergen} N. Bloembergen, Am. J. Phys. \textbf{35}, 989 (1967).


\end{thebibliography}
\end{document}